\documentclass[final,leqno]{siamltex}

\usepackage{amsmath,amsfonts,latexsym}

\def\01{\{0,1\}}
\newcommand{\ceil}[1]{\lceil{#1}\rceil}
\newcommand{\ket}[1]{|#1\rangle}
\newcommand{\bra}[1]{\langle#1|}
\newcommand{\inp}[2]{\langle{#1}|{#2}\rangle}

\newcommand{\DISJ}{\mbox{\rm DISJ}}

\newcommand{\EQ}{\mbox{\rm EQ}}
\newcommand{\NE}{\mbox{\rm NE}}

\newcommand{\AND}{\mbox{\rm AND}}
\newcommand{\OR}{\mbox{\rm OR}}

\newcommand{\PARITY}{\mbox{\rm PARITY}}

{\end{enumerate}\end{trivlist}\medskip}

\newcommand{\Dqc}{D}
\newcommand{\Rqc}{R}
\newcommand{\Nqc}{N}
\newcommand{\Qqc}{Q}

\renewcommand{\deg}{\text{{\it deg\/}}}
\newcommand{\ndeg}{\text{{\it ndeg\/}}}
\newcommand{\nrank}{\mathop{n\text{{\it rank}}}} 
\newcommand{\bs}{\text{{\it bs\/}}}
\newcommand{\Cov}{\mathop{\text{{\it Cov\/}}}}
\newcommand{\NQqc}{\text{{\it NQ\/}}}
\newcommand{\Dcc}{\text{{\it Dcc\/}}}
\newcommand{\Ncc}{\text{{\it Ncc\/}}}
\newcommand{\Qcc}{\text{{\it Qcc\/}}}
\newcommand{\NQcc}{\text{{\it NQcc\/}}}

\renewcommand{\ldots}{\dotsc}

\title{Nondeterministic Quantum Query and Communication Complexities\thanks{Received by the editors May 8, 2002; 
accepted for publication (in revised form) December~12, 2002;
	published electronically April 17, 2003\@.
	This paper combines results from the conference papers \cite{wolf:ndetq,hoyer&wolf:disjeq} with some new results.
\URL sicomp/32-3/40734.html}}

\author{Ronald de Wolf\thanks{CWI, 
	Kruislaan 413, 1098 SJ Amsterdam, The Netherlands
	(rdewolf@cwi.nl).
	This author was partially supported by the EU fifth framework project QAIP, IST--1999--11234.
	Part of this paper was written when the author was a postdoc at UC Berkeley, 
	supported by Talent grant S~\mbox{62--565} from the Netherlands Organization for Scientific Research (NWO)\@.}}

\begin{document}

\maketitle
\vspace{-1.2in}
\slugger{sicomp}{2003}{32}{3}{681--699}
\vspace{.9in}

\setcounter{page}{681}

\begin{abstract}
We study nondeterministic quantum algorithms for Boolean functions~$f$.
Such algorithms have positive acceptance probability on input~$x$
iff $f(x)=1$.
In the setting of query complexity, we show that the nondeterministic
quantum complexity of a Boolean function is equal to its ``nondeterministic
polynomial'' degree. We also prove a quantum-vs.-classical gap of
1~vs.~$n$ for nondeterministic query complexity for a total function.
In the setting of communication complexity, we show that the nondeterministic
quantum complexity of a two-party function is equal to the logarithm of
the rank of a nondeterministic version of the communication matrix.
This implies that the quantum communication complexities of the equality
and disjointness functions are $n+1$ if we do not allow any error probability.
We also exhibit a total function in which the nondeterministic quantum
communication complexity is exponentially smaller than its
classical counterpart.
\end{abstract}


\begin{AM}
68Q10
\end{AM}

\begin{PII}
S0097539702407345
\end{PII}

\pagestyle{myheadings}
\thispagestyle{plain}
\markboth{RONALD DE WOLF}{NONDETERMINISTIC QUANTUM COMPLEXITY}

\section{Introduction}

\subsection{Motivation}

In classical computing, {\it nondeterministic\/} computation has
a prominent place in many different models and for many good reasons.
For example, in Turing machine complexity, the study of nondeterminism
leads naturally to the class of NP-complete problems, which contains some
of the most important and practically relevant computer science
problems---as well as some of the hardest theoretical open questions.
In fields like query complexity and communication complexity,
there is a tight relation between deterministic complexity and
nondeterministic complexity, but it is often much easier
to analyze upper and lower bounds for the latter than for the former.

Suppose we want to compute a Boolean function~$f$
in some algorithmic setting, such as that of Turing machines,
decision trees, or communication protocols.
Consider the following two ways of viewing a nondeterministic algorithm.
The first and most common way is to think of it as a ``certificate verifier'':
a deterministic algorithm~$A$ that receives, apart from the input~$x$,
a ``certificate''~$y$ whose validity it needs to verify.
For all inputs~$x$, if $f(x)=1$, then there is a certificate~$y$ 
such that $A(x,y)=1$; if $f(x)=0$, then $A(x,y)=0$ for all~$y$.
Second, we may view $A$ as a {\it randomized\/} algorithm whose
acceptance probability is positive if $f(x)=1$ and whose
acceptance probability is zero if $f(x)=0$.
It is easy to see that these two views are equivalent in the classical case.
To turn an algorithm~$A$ of the first kind into one of the second kind,
we can just guess a certificate~$y$ at random and output~$A(x,y)$.
This will have positive acceptance probability iff $f(x)=1$.
For the other direction, we can consider the sequence of coin
flips used by an algorithm of the second kind as a certificate.
Clearly, there will be a certificate leading to output~1 iff $f(x)=1$,
which gives us an algorithm of the first kind.

Both views may be generalized to the quantum case, yielding {\it three\/}
potential definitions of nondeterministic quantum algorithms,
possibly nonequivalent.
The quantum algorithm may be required to output the right answer~$f(x)$
when given an appropriate certificate, which we can take
to be either quantum or classical. Or, third, the quantum algorithm may
be required to have positive acceptance  probability iff $f(x)=1$.
An example is given by two alternative definitions of quantum nondeterminism
in the case of quantum Turing machine complexity.
Kitaev defines the class
``bounded-error quantum-NP'' (BNQP) as the set of languages accepted
by polynomial-time bounded-error quantum algorithms that are given
a polynomial-size quantum certificate
(e.g., \cite{kitaev:qnp,kitaev&watrous:qip} and \cite[Chapter~14]{ksv:qc}).
On the other hand, Adleman, Demarrais, and Huang~\cite{adh:qcomputability} and
Fenner et al.~\cite{fghp:nqp} define quantum-NP
as the set of languages~$L$ for which there is
a polynomial-time quantum algorithm whose acceptance
probability is positive iff $x\in L$. This quantum class was shown to be equal to
the classical counting class \mbox{co-C$_=$P} \cite{fghp:nqp,yamakami&yao:c=p}
using tools from Fortnow and Rogers~\cite{fortnow&rogers:limitations}.

In this paper, we adopt the latter view: a nondeterministic quantum
algorithm for~$f$ is defined to be a quantum algorithm that outputs 1 with
positive probability if $f(x)=1$ and that always outputs 0 if $f(x)=0$.
This definition contrasts with the more traditional view of
classical determinism as ``certificate verification.''
The motivation for our choice of definition of quantum nondeterminism
is twofold.  First, in the appendix, we show that this definition
is strictly more powerful than the other two possible definitions
in the sense of being able to simulate the other definitions efficiently,
while the reverse is not true. Second,
it turns out that this definition lends itself to very crisp results.
Rather than in the quantum Turing machine setting of Kitaev, Adleman, etc.,
we study the complexity of nondeterministic algorithms in the
query complexity and communication complexity settings.
Our main results are exact characterizations of these nondeterministic quantum
complexities in algebraic terms and large gaps between quantum and
classical complexities in both settings. Our algebraic
characterizations can be extended to nontotal functions in the obvious way,
but we will stick to total functions in our presentation.

\subsection{Query complexity}

We first consider the model of query complexity,
also known as decision tree complexity or black box complexity.
Here the goal is to compute some function $f:\01^n\rightarrow\01$,
making as few queries to input bits as possible. Most existing quantum
algorithms can naturally be expressed in this model and
achieve provable speed-ups over the best classical algorithms.
Examples can be found, e.g., in \cite{deutsch&jozsa,simon:power,grover:search,brassard&hoyer:simon,bht:collision,bht:counting}
and also include the order-finding problem on which Shor's
celebrated factoring algorithm is based~\cite{shor:factoring}.

Let $\Dqc(f)$ and~$\Qqc_E(f)$ denote the query complexities
of optimal deterministic and quantum algorithms that compute
$f$ exactly.
Let $\deg(f)$ denote the minimal degree among all
multilinear polynomials that represent $f$.
(A polynomial~$p$ represents $f$ if $f(x)=p(x)$ for all $x\in\01^n$.)
The following relations are known. The first inequality is due to
Beals et al.~\cite{bbcmw:polynomials}, the second inequality
is obvious, and the last is due to Nisan and Smolensky---unpublished,
but described in the survey paper~\cite{buhrman&wolf:dectreesurvey}.
\[
\frac{\deg(f)}{2}\leq \Qqc_E(f)\leq \Dqc(f)\leq O(\deg(f)^4).
\]
Thus $\deg(f)$, $\Qqc_E(f)$, and~$\Dqc(f)$ are polynomially related for
all total $f$. (The situation is very different for partial
$f$ \cite{deutsch&jozsa,simon:power,shor:factoring,bcw:sharp}.)
Nisan and Szegedy~\cite{nisan&szegedy:degree} exhibit
a function with a large gap between $\Dqc(f)=n$ and
$\deg(f)=n^{0.6\ldots}$, but no function is known where
$\Qqc_E(f)$ is significantly larger than~$\deg(f)$, and it may
in fact be true that $\Qqc_E(f)$ and~$\deg(f)$ are linearly related.
In section~\ref{secndetquery}, we show that the {\it nondeterministic\/}
versions of $\Qqc_E(f)$~and~$\deg(f)$ are in fact {\it equal\/}:
\[
\NQqc(f)=\ndeg(f).
\]
Here $\NQqc(f)$ denotes the query complexity of an optimal
nondeterministic quantum algorithm for~$f$, which has nonzero
acceptance probability iff $f(x)=1$. The nondeterministic degree~$\ndeg(f)$ is the minimal degree of a so-called 
{\it nondeterministic\/}
polynomial for~$f$, which is required to be nonzero iff $f(x)=1$.
A note on terminology: the name ``nondeterministic polynomial''
is based only on analogy with the acceptance probability of
a nondeterministic algorithm.
This name is less than ideal, since such polynomials have little to
do with the traditional view of nondeterminism as certificate verification.
Nevertheless, we use this name because any alternatives that we could think
of were worse (too verbose or confusing).

Apart from the algebraic characterization of the nondeterministic
quantum query complexity $\NQqc(f)$, we also show that $\NQqc(f)$
may be much smaller than its classical analogue~$\Nqc(f)$:
we exhibit an $f$ where $\NQqc(f)=1$ and $\Nqc(f)=n$, which is the biggest
possible gap allowed by this model.
Accordingly, while the case of exact (or, for that matter, bounded-error)
computation allows at most polynomial quantum-classical query complexity
gaps for total functions,
the nondeterministic case allows {\it unbounded\/} gaps.

\subsection{Communication complexity}

In the case of communication complexity, the goal is for two
distributed parties, Alice and Bob, to compute some function
$f:\01^n\times\01^n\rightarrow\01$. Alice receives an $x\in\nobreak\01^n$, and
Bob receives a $y\in\nobreak\01^n$, and they want to compute $f(x,y)$,
exchanging as few bits of communication as possible.
This model was introduced by Yao~\cite{yao:distributive} and is fairly
well understood for the case in which Alice and Bob are classical players
exchanging classical bits~\cite{kushilevitz&nisan:cc}.
Much less is known about {\it quantum\/} communication complexity,
where Alice and Bob have a quantum computer and can exchange qubits.
This was first studied by Yao~\cite{yao:qcircuit}, and it was shown
later that quantum communication complexity can be significantly
smaller than classical communication
complexity \cite{cleve&buhrman:subs,BuhrmanCleveWigderson98,astv:qsampling,raz:qcc,bcww:fp}.

Let $\Dcc(f)$ and~$\Qcc_E(f)$ denote the communication required for optimal
deterministic classical and exact quantum protocols for computing $f$,
respectively.\footnote{The notation~$D(f)$ is used for deterministic
complexity in decision tree complexity as well as in communication
complexity. To avoid confusion, we will consistently add ``cc'' to indicate
communication complexity.}
Here we assume Alice and Bob do not share any randomness or prior entanglement.
Let ${\mbox{\it rank}}(f)$ be the rank of the $2^n\times 2^n$ communication matrix~$M_f$,
which is defined by $M_f(x,y)=f(x,y)$. The following relations are known:
\[
\frac{\log {\mbox{\it rank}}(f)}{2}\leq \Qcc_E(f)\leq \Dcc(f).
\]
The first inequality follows from work of Kremer~\cite{kremer:thesis}
and Yao~\cite{yao:qcircuit}, as first noted by
Buhrman, Cleve, and Wigderson~\cite{BuhrmanCleveWigderson98}.
(In~\cite{buhrman&wolf:qcclower} it is shown
that this lower bound also holds if the quantum protocol can make use of
unlimited prior entanglement between Alice and Bob.)
It is an open question whether $\Dcc(f)$ can in turn
be upper bounded by some polynomial in $\log {\mbox{\it rank}}(f)$.
The conjecture that it can is known as the {\it log-rank conjecture}.
If this conjecture holds, then $\Dcc(f)$ and~$\Qcc_E(f)$
are polynomially related for all total $f$ (which may well be true).
It is known that $\log {\mbox{\it rank}}(f)$ and~$\Dcc(f)$ are not linearly
related~\cite{nisan&wigderson:rank}.
In section~\ref{secndetcc}, we show that the {\it nondeterministic\/}
version of $\log {\mbox{\it rank}}(f)$ in fact fully determines the nondeterministic
version of~$\Qcc_E(f)$:
\[
\NQcc(f)=\ceil{\log \nrank(f)}+1.
\]
Here $\nrank(f)$ denotes the minimal rank of a matrix whose $(x,y)$-entry
is nonzero iff $f(x,y)=1$.
Thus we can characterize the nondeterministic quantum communication
complexity fully by the logarithm of the rank of its nondeterministic matrix.
As far as we know, only two other log-rank-style characterizations of certain
variants of communication complexity are known: the communication complexity of
quantum sampling due to Ambainis et al.~\cite{astv:qsampling} and the
so-called modular communication complexity due to Meinel and
Waack~\cite{meinel&waack:logrankmod}.

Equality and disjointness both have nondeterministic rank~$2^n$,
so their nondeterministic complexities are maximal:
$\NQcc(\EQ)=\NQcc(\DISJ)=n+1$.
Since $\NQcc(f)$ lower bounds $\Qcc_E(f)$, we also obtain optimal bounds
for the exact quantum communication complexity of
equality and disjointness.  In particular, for the equality function,
we get $\Qcc_E(\EQ)=n+1$, which answers a question posed by
Gilles Brassard in a personal communication \cite{new}. 
Surprisingly, no proof of this fact seems to be known that avoids
our detour via nondeterministic computation.
Thus our methods also give new lower bounds for
regular quantum communication complexity.

Finally, analogous to the query complexity case,
we also show an exponential gap between quantum and classical nondeterministic
communication complexity: we exhibit an $f$ where $\NQcc(f)\leq\log(n+1)+1$
and $\Ncc(f)\in\Omega(n)$. Massar et al.~\cite{mbcc:siment} earlier found
another gap that is unbounded, yet in some sense smaller:
$\NQcc(\NE)=2$ versus $\Ncc(\NE)=\log n+1$,
where $\NE$ is the nonequality function.

\section{Nondeterministic quantum query complexity}\label{secndetquery}

\subsection{Functions and polynomials}\label{ssecfpoly}

For $x\in\01^n$, we use $|x|$ for the Hamming weight (number of 1's) 
of~$x$, and $x_i$ for its $i$th bit, $i\in [n]=\{1,\ldots,n\}$.
We use $\vec{0}$ for a string of $n$~zeros.
If $B\subseteq[n]$ is a set of (indices of) variables,
then $x^B$ denotes the input obtained from~$x$
by complementing all variables in~$B$.
If $x,y\in\01^n$, then $x\wedge y$ denotes the $n$-bit string obtained
by bitwise ANDing $x$ and~$y$.
Let $f:\01^n\rightarrow\01$ be a total Boolean function.
For example, $\OR(x)=1$ iff $|x|\geq 1$, $\AND(x)=1$ iff $|x|=n$,
$\PARITY(x)=1$ iff $|x|$ is odd.
We use $\overline{f}$ for the function $1-f$.

For $b\in\01$, a {\it $b$-certificate\/} for~$f$ is an assignment
$C:S\rightarrow\01$ to some set~$S$ of variables,
such that $f(x)=b$ whenever $x$ is consistent with~$C$.
The {\it size\/} of~$C$ is $|S|$.
The {\it certificate complexity}~$C_x(f)$ of~$f$ on input~$x$ is the
minimal size of an $f(x)$-certificate that is consistent with~$x$.
We define the 1-certificate complexity of~$f$ as
$C^
{(1)}(f)=\max_{x:f(x)=1}C_x(f)$. We define $C^{(0)}(f)$ similarly.
For example, $C^{(1)}(\OR)=1$ and $C^{(0)}(\OR)=n$,
but $C^{(1)}(\overline{\OR})=n$ and $C^{(0)}(\overline{\OR})=1$.

An $n$-variate {\it multilinear polynomial\/}
is a function $p:\mathbb{C}^n\rightarrow\mathbb{C}$ that can be written
\[
p(x)=\sum_{S\subseteq[n]}a_SX_S.
\]
Here $S$ ranges over all sets of indices of variables,
$a_S$ is a complex number, and the monomial~$X_S$ is the product
$\Pi_{i\in S}x_i$ of all variables in~$S$.
The {\it degree}~$\deg(p)$ of~$p$ is the degree of a largest monomial with
nonzero coefficient.
It is well known that every total Boolean $f$ has a unique polynomial~$p$ 
such that $p(x)=f(x)$ for all $x\in\01^n$.
Let $\deg(f)$ be the degree of this polynomial, which is at most~$n$.
For example, $\OR(x_1,x_2)=x_1+x_2-x_1x_2$, which has degree~2.
Every multilinear polynomial $p=\sum_S a_SX_S$ can also be written out
uniquely in the so-called {\it Fourier basis\/}:
\[
p(x)=\sum_S c_S(-1)^{x\cdot S}.
\]
Again $S$ ranges over all sets of indices of variables (we often identify
a set~$S$ with its characteristic $n$-bit vector), $c_S$ is a complex number,
and $x\cdot S$ denotes the inner product of the $n$-bit strings $x$ and~$S$, or,
equivalently, $x\cdot S=|x\wedge S|=\sum_{i\in S}x_i$.
It is easy to see that $\deg(p)=\max\{|S|\mid c_S\neq 0\}$.
For example, $\OR(x_1,x_2)=\frac{3}{4}-\frac{1}{4}(-1)^{x_1}-\frac{1}{4}(-1)^{x_2}-\frac{1}{4}(-1)^{x_1+x_2}$ in the Fourier basis.
We refer to \cite{beigel:poly,nisan&szegedy:degree,buhrman&wolf:dectreesurvey}
for more details about polynomial representations of Boolean functions.

We introduce the notion of a {\it nondeterministic polynomial\/} for~$f$.
This is a polynomial~$p$ such that $p(x)\neq 0$ iff $f(x)=1$.
Let the {\it nondeterministic degree\/} of~$f$, denoted $\ndeg(f)$,
be the minimum degree among all nondeterministic polynomials~$p$ for~$f$.
For example, $p(x)=\sum_{i=1}^n x_i$ is a nondeterministic polynomial
for $\OR$; hence $\ndeg(\OR)=1$.

We mention some upper and lower bounds for $\ndeg(f)$.
Let $f$ be a nonconstant symmetric function
(i.e., $f(x)$ depends only on~$|x|$). Suppose $f$ achieves value~0
on the $z$~Hamming weights, $k_1,\ldots,k_z$.
Since $|x|=\sum_ix_i$, it is easy to see that
$(|x|-\nobreak k_1)(|x|-\nobreak k_2)\cdots(|x|-\nobreak k_z)$
is a nondeterministic polynomial for~$f$; hence $\ndeg(f)\leq z$.
This upper bound is tight for $\AND$ (see below) but not for $\PARITY$.
For example, $p(x_1,x_2)=\nobreak x_1-\nobreak x_2$ is a degree-1 nondeterministic
polynomial for $\PARITY$ on two variables: it assumes value~0 on $x$-weights 0 and~2 and $\pm 1$ on weight~1.
By squaring~$p(x)$ and then using standard symmetrization techniques
(as used, for instance,
in \cite{minsky&papert:perceptrons,nisan&szegedy:degree,bbcmw:polynomials}),
we can also show the general lower bound $\ndeg(f)\geq z/2$ for symmetric~$f$.
Furthermore, it is easy to show that $\ndeg(f)\leq C^{(1)}(f)$ for every~$f$.
(Take a polynomial that is the ``sum'' over all 1-certificates for~$f$.)

Finally, we mention a general lower bound on $\ndeg(f)$.
Let $\Pr[{p\neq\nobreak 0}]=\break 
|\{x\in\nobreak\01^n\mid\nobreak p(x)\neq\nobreak 0\}|/2^n$ denote the probability
that a random Boolean input~$x$ makes a function~$p$ nonzero.
A lemma of Schwartz~\cite{schwartz:probabilistic}
(see also \cite[section~2.2]{nisan&szegedy:degree}) states that
if $p$ is a nonconstant multilinear polynomial of degree~$d$,
then $\Pr[p\neq 0]\geq 2^{-d}$, and hence $d\geq\log(1/\Pr[p\neq 0])$.
Since a nondeterministic polynomial~$p$ for~$f$ is nonzero iff
$f(x)=1$, it follows that
\[
\ndeg(f)\geq\log(1/\Pr[f\neq 0])=\log(1/\Pr[f=1]).
\]
Accordingly, functions with a very small fraction of 1-inputs will have
high nondeterministic degree. For instance, $\Pr[\AND=1]=2^{-n}$,
so $\ndeg(\AND)=n$.

\subsection{Quantum computing}

We assume familiarity with classical computation and briefly
sketch the setting of quantum computation
(see, e.g.,~\cite{nielsen&chuang:qc} for more details).
An {\it $m$-qubit state\/} is a linear combination of all classical $m$-bit states
\[
\ket{\phi}=\sum_{i\in\01^m}\alpha_i\ket{i},
\]
where $\ket{i}$ denotes the basis state~$i$ (a classical $m$-bit string)
and $\alpha_i$ is a complex number that is called the
{\it amplitude\/} of~$\ket{i}$. We require $\sum_i|\alpha_i|^2=1$.
Viewing $\ket{\phi}$ as a $2^m$-dimensional column vector, we use
$\bra{\phi}$ for the row vector that is the conjugate transpose of~$\ket{\phi}$. 
Note that the inner product $\bra{i}\ket{j}=\inp{i}{j}$
is 1 if $i=j$ and 0 if $i\neq j$.
When we observe $\ket{\phi}$, we will see $\ket{i}$ with probability
$|\inp{i}{\phi}|^2=|\alpha_i|^2$,
and the state will collapse to the observed $\ket{i}$.
A quantum operation which is not an observation
corresponds to a unitary (i.e., norm-preserving) transformation~$U$
on the $2^m$-dimensional vector of amplitudes.

\subsection{Query complexity}

Suppose we want to compute some function $f:\break\01^n\rightarrow\nobreak\01$.
For input $x\in\01^n$, a {\it query\/} corresponds to
the unitary transformation~$O$ that maps
$\ket{i,b,z}\rightarrow\ket{i,b\oplus x_i,z}$.
Here $i\in[n]$ and $b\in\01$;
the $z$-part corresponds to the workspace, which is not affected by the query.
We assume that the input can be accessed only via such queries.
A $T$-query quantum algorithm has the form $A=U_TOU_{T-1}\cdots OU_1OU_0$,
where the $U_k$ are fixed unitary transformations, independent of the input~$x$.
This $A$ depends on~$x$ via the $T$~applications of~$O$.
We sometimes write $A_x$ to emphasize this.
The algorithm starts in initial state~$\ket{\vec{0}}$, and its {\it output\/}
is the bit obtained from observing the leftmost qubit of the
final superposition~$A\ket{\vec{0}}$.
The {\it acceptance probability\/} of~$A$ (on input~$x$)
is its probability of outputting~1 (on~$x$).

We will consider classical and quantum algorithms and will count only
the number of queries these algorithms make on a worst-case input.
Let $\Dqc(f)$ and~$\Qqc_E(f)$ be the query complexities of optimal deterministic
classical and exact quantum algorithms for computing $f$, respectively.
$\Dqc(f)$ is also known as the decision tree complexity of~$f$.
Similarly we can define $\Rqc_2(f)$ and~$\Qqc_2(f)$ to be the
query complexity of~$f$ for bounded-error classical and quantum
algorithms, respectively.
Quantum query complexity and its relation to classical complexity has been
well studied in recent years; see, for example, \cite{bbcmw:polynomials,ambainis:lowerbounds,buhrman&wolf:dectreesurvey}.

We define a {\it nondeterministic algorithm\/} for~$f$ to be
an algorithm that has positive acceptance probability on input~$x$ iff $f(x)=1$.
Let $\Nqc(f)$ and~$\NQqc(f)$ be the query complexities of optimal
nondeterministic classical and quantum algorithms for~$f$, respectively.
It is easy to show that the 1-certificate complexity fully characterizes
the classical nondeterministic complexity of~$f$.

\begin{proposition}
$\Nqc(f)=C^{(1)}(f)$.
\end{proposition}

\begin{proof}
A classical algorithm that
guesses a 1-certificate, queries its variables, and outputs 1 iff
the certificate holds is a nondeterministic algorithm for~$f$.
Hence $\Nqc(f)\leq C^{(1)}(f)$.

A nondeterministic algorithm for~$f$ can only output 1 if the outcomes
of the queries that it has made force the function to~1.
Hence, if $x$ is an input where all 1-certificates have size at least~$C^{(1)}(f)$, 
then the algorithm will have to query at least $C^{(1)}(f)$~variables before it can output 1 (which it must do on some runs).
Hence $\Nqc(f)\geq C^{(1)}(f)$.\qquad\end{proof}

\subsection{Algebraic characterization}

Here we show that $\NQqc(f)$ is equal to $\ndeg(f)$,
using the following result from~\cite{bbcmw:polynomials}.

\begin{lemma}[see \cite{bbcmw:polynomials}]\label{lem2Tpoly}
The amplitudes of the basis states in the final superposition of
a $T$-query quantum algorithm can be written as multilinear
complex-valued polynomials of degree\/ $\leq T$ in the $n$~$x_i$-variables.
Therefore, the acceptance probability of the algorithm (which is the sum
of squares of some of those amplitudes) can be written as an $n$-variate
multilinear polynomial~$P(x)$ of degree\/ $\leq 2T$.
\end{lemma}

Note that the acceptance probability of a nondeterministic quantum
algorithm is actually a nondeterministic polynomial for~$f$,
since it is positive iff $f(x)=1$. By Lemma~\ref{lem2Tpoly},
this polynomial will have degree at most twice the number of queries
of the algorithm, which immediately implies $\ndeg(f)/2\leq \NQqc(f)$.
Below we will show how we can get rid of the factor~$1/2$ in this
lower bound, improving it to $\ndeg(f)\leq \NQcc(f)$.
We show that this lower bound is in fact optimal by deriving
a nondeterministic algorithm from a nondeterministic polynomial.
This derivation uses a trick similar to the one used in~\cite{fghp:nqp}
to show that $\mbox{co-C$_=$P} \subseteq \mbox{quantum-NP}$.

\begin{theorem}\label{thNQ=ndeg}
$\NQqc(f)=\ndeg(f)$.
\end{theorem}

\begin{proof}
{\it Upper bound}.
Let $p(x)$ be a nondeterministic
polynomial for~$f$ of degree $d=\ndeg(f)$.
Recall that $x\cdot S$ denotes $|x\wedge S|$, identifying
$S\subseteq[n]$ with its characteristic $n$-bit vector.
We write $p$ in the Fourier basis:
\[
p(x)=\sum_S c_S(-1)^{x\cdot S}.
\]
Since $\deg(p)=\max\{|S|\mid c_S\neq 0\}$, we have that
$c_S\neq 0$ only if $|S|\leq d$.

We can construct a unitary transformation~$F$ that uses $d$
queries to~$x$ and maps $\ket{S}\rightarrow (-1)^{x\cdot S}\ket{S}$
whenever $|S|\leq d$. Informally, this transformation does
a controlled parity-computation: it computes $|x\cdot S|\pmod 2$
using $|S|/2$~queries \cite{bbcmw:polynomials,fggs:parity},
then adds a phase~``$-1$'' if that answer is 1,
and then reverses the computation to clean up the workspace and the answer
at the cost of another $|S|/2$~queries. (If $|S|$ is odd, then
one variable is treated separately, still using $|S|$~queries in total.)

Now consider the following quantum algorithm:
\begin{enumerate}
\item Start with $c\sum_S c_S\ket{S}$
(an $n$-qubit state, where $c=1/\sqrt{\sum_S |c_S|^2}$ is a normalizing constant).
\item Apply $F$ to the state.
\item Apply a Hadamard transform~$H$ to each qubit.
\item Measure the final state, and output 1 if the outcome is the
all-zero state~$\ket{\vec{0}}$, and output 0 otherwise.
\end{enumerate}
The state after step~2 is $c\sum_S c_S(-1)^{x\cdot S}\ket{S}$.
Note that the sum of the amplitudes in this state is $c\cdot p(x)$,
which is nonzero iff $f(x)=1$.
The Hadamard transform in step~3 gives us this sum as amplitude
of the $\ket{\vec{0}}$-state, with a normalizing factor of~$1/\sqrt{2^n}$.
Accordingly, the probability of observing $\ket{\vec{0}}$ at the end is
\begin{eqnarray*} 
P(x) & = & \left|\bra{\vec{0}}H^{\otimes n}Fc\sum_S c_S\ket{S}\right|^2\\
     & = & \frac{c^2}{2^n}\left|\sum_{S'}\bra{S'}\sum_S c_S(-1)^{x\cdot S}\ket{S}\right|^2\\
     & = & \frac{c^2}{2^n}\left|\sum_{S}c_S(-1)^{x\cdot S}\right|^2\\
     & = & \frac{c^2p(x)^2}{2^n}.
\end{eqnarray*}
Since $p(x)$ is nonzero iff $f(x)=1$, $P(x)$ will be positive iff $f(x)=1$.
Hence we have a nondeterministic quantum algorithm for~$f$
with $d=\ndeg(f)$ queries.

{\it Lower bound}.
Let $T=\NQqc(f)$, and consider a $T$-query nondeterministic quantum algorithm for~$f$.
By Lemma~\ref{lem2Tpoly}, the amplitudes~$\alpha_i$ in the final state,
\[
\ket{\phi^x}=\sum_i\alpha_i(x)\ket{i},
\]
on input~$x$ are $n$-variate polynomials of~$x$ of degree $\leq T$.
We use the probabilistic method~\cite{alon&spencer:probmethod}
to show that some linear combination of these polynomials is
a nondeterministic polynomial for~$f$, thus avoiding losing
the factor~$1/2$ mentioned after Lemma~\ref{lem2Tpoly}.

Let $S$ be the set of basis states having a 1 as leftmost bit
(observing such a state will lead the algorithm to output 1).
Since the algorithm is nondeterministic, we have the following properties:
\begin{quote}
If $f(x)=0$, then $\alpha_i(x)=0$ for all $i\in S$.\\
If $f(x)=1$, then $\alpha_i(x)\neq 0$ for at least one $i\in S$.
\end{quote}
Let $I$ be an arbitrary set of more than $2^n$~numbers.
For each $i\in S$, pick a coefficient~$c_i$ uniformly at random from~$I$,
and define $p(x)=\sum_{i\in S}c_i\alpha_i(x)$.
By the first property, we have $p(x)=0$ whenever $f(x)=0$.
Now consider an $x$ for which $f(x)=1$, and let $k\in S$ satisfy $a=\alpha_k(x)\neq 0$.
Such a $k$ must exist by the second property.
We want to show that the event $p(x)=0$ happens only with very small probability
(probability taken over the random choices of the $c_i$).
In order to do this, we fix the random choices~$c_i$ for all $i\neq k$
and view $p(x)=ac_k+b$ as a linear function in the only not-yet-chosen coefficient~$c_k$.
Since $a\neq 0$, at most one out of $|I|>2^n$ many possible choices
of~$c_k$ can make $p(x)=0$, so
\[
\Pr[p(x)=0]<2^{-n}.
\]
However, then, by the union bound we have
\begin{multline*}
\Pr\left[\mbox{there is an }x
           \in f^{-1}(1)\mbox{ for which }p(x)=0\right] \\
\leq \sum_{x\in f^{-1}(1)}\Pr[p(x)=0]< 2^n\cdot 2^{-n}
=1.
\end{multline*}
This probability is strictly less than~1,
which shows that there exists a way of setting the coefficients~$c_i$
that satisfies $p(x)\neq 0$ for all $x\in f^{-1}(1)$,
thus making $p$ a nondeterministic polynomial for~$f$.
Since $p$ is a sum of polynomials of degree $\leq T$,
it follows that $\ndeg(f)\leq \deg(p)\leq T=\NQqc(f)$.\qquad\end{proof}

\subsection{Quantum-classical separation}\label{ssecnqquerysep}

What is the biggest possible gap between quantum and classical
nondeterministic query complexity?
Consider the total Boolean function $f:\01^n\rightarrow\01$ defined by
\[
f(x)=1 \ {\mbox {\rm iff}} \ |x|\neq 1.
\]
It is easy to see that $\Nqc(f)=C^{(1)}(f)=C^{(0)}(f)=n$.
On the other hand, the following is a degree-1 nondeterministic polynomial for~$f$:
\begin{equation}\label{eqndetpoly}
p(x)=\left(\sum_{i=1}^n x_i\right)-1=\frac{n}{2}-1-\frac{1}{2}\sum_{i=1}^n(-1)^{x_i}.
\end{equation}
Thus we have that $\NQqc(f)=\ndeg(f)=1$.
Explicitly, the 1-query algorithm that we get from
the proof is as follows:
\begin{enumerate}
\item Start with $c\,((n/2-1)\ket{\vec{0}}-(1/2)\sum_i\ket{e_i})$,
where $c=1/\sqrt{n^2/4-3n/4+1}$ and $\ket{e_i}$ has a 1 only at the $i$th bit.
\item Using one query, we can map $\ket{e_i}\rightarrow(-1)^{x_i}\ket{e_i}$.
\item Applying a Hadamard transform
turns the amplitude of $\ket{\vec{0}}$ into
$\alpha_{\vec{0}}=\break\frac{c}{\sqrt{2^n}}\left((n/2-1)-\sum_i(-1)^{x_i}/2\right)=
cp(x)/\sqrt{2^n}$.
\item Hence the probability of observing $\ket{\vec{0}}$
at the end is $\alpha_{\vec{0}}^2=c^2p(x)^2/2^n$.
\end{enumerate}
For the complement of~$f$, we can easily show
$\NQqc(\overline{f})=\ndeg(\overline{f})\geq n-1$
(the ``$-1$'' is tight for~$n=2$; witness $p(x)=x_1-x_2$).
In sum, we have the following theorem.

\begin{theorem}
For the above $f$, we have $\NQqc(f)=1$, $\NQqc(\overline{f})\geq n-1$,
and $\Nqc(f)=\Nqc(\overline{f})=n$.
\end{theorem}

\subsection{Relation to some other complexity measures}\label{ssecembedndeg}

Many relations are known between all sorts of complexity measures
of Boolean functions, such as polynomial degree, certificate complexity,
various classical and quantum decision tree complexities, etc.
A survey may be found in~\cite{buhrman&wolf:dectreesurvey}.
In this subsection, we will similarly embed $\ndeg(f)$ ($=\NQqc(f)$)
in this web of relations and give upper bounds on~$\Dqc(f)$ in terms
of $\ndeg(f)$, $C(f)$, and the {\it block sensitivity}~$\bs(f)$,
which is defined as follows. A set of (indices of)
variables $B\subseteq[n]$ is called a {\it sensitive block\/}
for~$f$ on input~$x$ if $f(x)\neq f(x^B)$;
$B$ is {\it minimal\/} if no $B'\subset B$ is sensitive.
The block sensitivity~$\bs_x(f)$ is the maximal number of disjoint
minimal sensitive blocks in~$x$,
and $\bs^{(b)}(f)=\max_{x\in f^{-1}(b)}\bs_x(f)$.

\begin{lemma}
If $f(x)=0$ and $B$ is a minimal sensitive block for~$f$ on~$x$,
then\/ $|B|\leq \ndeg(f)$.
\end{lemma}

\begin{proof}
Assume without loss of generality that $x=\vec{0}$.
Because $B$ is minimal, for every proper subset~$B'$ of~$B$,
we have $f(x)=f(x^{B'})=0$, but on the other hand $f(x^B)=1$.
Accordingly, if we fix all variables outside of~$B$ to zero,
then we obtain the AND-function of $|B|$~variables,
which requires nondeterministic degree~$|B|$.
Hence $|B|\leq \ndeg(f)$.\qquad\end{proof}

\begin{lemma}
$C^{(0)}(f)\leq \bs^{(0)}(f)\ndeg(f)$.
\end{lemma}

\begin{proof}
Consider any input~$x$.  As Nisan~\cite{nisan:pram&dt} proved,
the union of a maximal set of sensitive blocks forms
a certificate for that input (for otherwise there would be one
more sensitive block). If $f(x)=0$, then there can be at most
$\bs^{(0)}(f)$ disjoint sensitive blocks, and by the previous lemma
each block contains at most $\ndeg(f)$ variables.
Hence each 0-input contains a certificate of at most
$\bs^{(0)}(f)\ndeg(f)$ variables.\qquad\end{proof}

The following theorem improves upon an argument of Nisan and Smolensky,
described in~\cite{buhrman&wolf:dectreesurvey}.

\begin{theorem}
$\Dqc(f)\leq C^{(0)}(f)\ndeg(f)$.
\end{theorem}

\begin{proof}
Let $p$ be a nondeterministic polynomial for~$f$ of degree $d=\ndeg(f)$.
Note that if we take a 0-certificate $C:S\rightarrow\01$
and fix the $S$-variables accordingly,
then $p$ must reduce to the constant-0 polynomial.
This implies that $S$ intersects all degree-$d$ monomials of~$p$,
because a nonintersected degree-$d$ monomial would still be present in
the reduced polynomial, which would then not be constant-0.
Thus taking a minimal 0-certificate and querying its variables
reduces the degree of~$p$ by at least~1. Repeating this at most
$\ndeg(f)$ times, we reduce $p$ to a constant polynomial and know $f(x)$.
This algorithm takes at most $C^{(0)}(f)\ndeg(f)$ queries.\qquad\end{proof}

Combining this with the fact that $\bs^{(0)}(f)\leq 6\Qqc_2(f)^2$~\cite{bbcmw:polynomials},
we obtain the following.

\begin{corollary}
$\Dqc(f)\leq \bs^{(0)}(f)\ndeg(f)^2\leq 6\ \Qqc_2(f)^2\NQqc(f)^2$.
\end{corollary}

This corollary has the somewhat paradoxical consequence
that if the nondeterministic complexity~$\NQqc(f)$ is small,
then the bounded-error complexity~$\Qqc_2(f)$ must be large
(i.e., close to~$\Dqc(f)$).
For instance, if $\NQqc(f)=O(1)$, then $\Qqc_2(f)=\Omega(\sqrt{\Dqc(f)})$.
We hope that this result will help tighten the relation
$\Dqc(f)=O(\Qqc_2(f)^6)$ that was proved in~\cite{bbcmw:polynomials}.

\pagebreak
\section{Nondeterministic quantum communication complexity}\label{secndetcc}

\subsection{Communication complexity}

In the standard version of communication complexity, two parties
(Alice and Bob) want to compute some function $f:\01^n\times\nobreak\01^n\rightarrow\nobreak\01$.
For example, $\EQ(x,y)=1$ iff $x=y$, $\NE(x,y)=1$ iff $x\neq y$, and
$\DISJ(x,y)=1$ iff $|x\wedge y|=0$.
A {\it rectangle\/} is a subset $R=S\times T$ of the domain of~$f$.
$R$ is a {\it {\rm1}-rectangle\/} (for~$f$) if $f(x,y)=1$ for all $(x,y)\in R$.
A {\it {\rm1}-cover\/} for~$f$ is a set of 1-rectangles whose union contains
all 1-inputs of~$f$.
$\Cov^1(f)$ denotes the minimal size (i.e., minimal number of rectangles)
of a 1-cover for~$f$.
Similarly, we define 0-rectangles, 0-covers, and $\Cov^0(f)$.

The {\it communication matrix}~$M_f$ of~$f$ is the $2^n\times 2^n$ Boolean matrix
whose $(x,y)$-entry is $f(x,y)$, and ${\mbox{\it rank}}(f)$ denotes the rank of~$M_f$
over the field of complex numbers.
A $2^n\times 2^n$ matrix~$M$ is called a {\it nondeterministic communication
matrix\/} for~$f$ if it has the property that $M(x,y)\neq 0$ iff $f(x,y)=1$.
Thus $M$ is any matrix obtainable by replacing 1-entries in~$M_f$ by nonzero
complex numbers.
Let the {\it nondeterministic rank\/} of~$f$, denoted $\nrank(f)$,
be the minimum rank (over the complex field) among
all nondeterministic matrices~$M$ for~$f$.\footnote{This definition looks
somewhat similar to the definition of the
{\it Colin de Verdi\'ere parameter}~$\mu(G)$
of an undirected graph~$G$~\cite{hls:colin}.
For $G=(V,E)$ with $|V|=n$,
$\mu(G)$ is defined to be the maximal corank ($=n-{\mbox{\it rank}}$) among all real
symmetric $n\times n$ matrices~$M$ having the following three properties:
(1)~$M_{ij}<0$ if $(i,j)\in E$ and $M_{ij}=0$ if $i\neq j$ and $(i,j)\notin E$;
(2)~$M$ has exactly one negative eigenvalue of multiplicity~1;
(3)~there is no real symmetric matrix $X\neq 0$ such that $MX=0$
and $X_{ij}=0$ whenever $i=j$ or $M_{ij}\neq 0$.
Such a matrix~$M$ is a nondeterministic matrix for the communication
complexity problem $f:[n]\times[n]\rightarrow\01$ defined by $f(i,j)=1$ iff
$(i,j)\in E$, with the promise that the inputs $i$ and~$j$ are distinct.
However, the Colin de Verdi\`ere requirement
appears to be more stringent, since it constrains the nondeterministic matrix
further by properties (2) and~(3).}

We consider classical and quantum communication protocols
and count only the amount of communication (bits or qubits) that
these protocols make on a worst-case input.
For classical communication protocols, we refer to~\cite{kushilevitz&nisan:cc}.
Here we briefly define quantum communication protocols, referring to the
surveys \cite{tashma:qcc,buhrman:qccsurvey,klauck:qccsurvey,brassard:qcc,wolf:qccsurvey}
for more details.
The space in which the quantum protocol works consists of three parts:
Alice's part, the communication channel, and Bob's part. (We do not
write the dimensions of these spaces explicitly.)
Initially these three parts contain only 0-qubits,
\[
\ket{0}\ket{0}\ket{0}.
\]
We assume Alice starts the protocol.
She applies a unitary transformation~$U^A_1(x)$ to her private space and
part of the channel.  This corresponds to her initial computation and
her first message. The length of this message is the number of channel
qubits on which $U^A_1(x)$ acts. The total state is now
\[
(U^A_1(x)\otimes I^B)\ket{0}\ket{0}\ket{0},
\]
where $\otimes$ denotes tensor product, and $I^B$ denotes
the identity transformation on Bob's part.
Then Bob applies a unitary transformation $U^B_2(y)=V^B_2(y)S^B_2$
to his part and the channel. First, the operation~$S^B_2$
``reads'' Alice's message by swapping the contents of the channel
with some fresh $\ket{0}$-qubits in Bob's private space.
After this, the unitary~$V^B_2(y)$ is applied to Bob's private space
and part of the channel. This corresponds to Bob's private computation
and his putting a message to Alice on the channel. The length of this
new message is the number of channel-qubits on which $V^B_2(y)$ acts.
This process goes back and forth for some $k$~messages,
so the final state of the protocol on input~$(x,y)$ will be
(in case Alice goes last as well)
\[
(U^A_k(x)\otimes I^B)(I^A\otimes U^B_{k-1}(y))\cdots
(I^A\otimes U^B_2(y))(U^A_1(x)\otimes I^B)\ket{0}\ket{0}\ket{0}.
\]
The total {\it cost\/} of the protocol is the total length
of all messages sent, on a worst-case input~$(x,y)$.
For technical convenience, we assume that at the end of the protocol the
output bit is the first qubit on the channel.
Thus the acceptance probability~$P(x,y)$ of the protocol is the probability that
a measurement of the final state gives a ``1'' in the first channel-qubit.
Note that we do not allow intermediate measurements during the protocol.
This is without loss of generality; it is well known that such measurements
can be postponed until the end of the protocol at no extra communication cost.

Let $\Dcc(f)$ and~$\Qcc_E(f)$ be the communication complexities of optimal deterministic
classical and quantum protocols for computing $f$, respectively.
A {\it nondeterministic protocol\/} for~$f$ is a protocol that has positive
acceptance probability on input~$(x,y)$ iff $f(x,y)=1$.
Let $\Ncc(f)$ and~$\NQcc(f)$ be the communication complexities of optimal
nondeterministic classical and quantum protocols for~$f$, respectively.
Our $\Ncc(f)$ is called $N^1(f)$ in~\cite{kushilevitz&nisan:cc}.

It is not hard to show that $\Ncc(f)=\ceil{\log \Cov^1(f)}+1$,
where the ``$+1$'' is due to the fact that we want Alice and Bob both to know
the output at the end of the protocol.

\subsection{Algebraic characterization}

Here we characterize $\NQcc(f)$ in terms of $\nrank(f)$.
We use the following lemma.
It was stated without proof by Yao~\cite{yao:qcircuit}
and in more detail by Kremer~\cite{kremer:thesis} and is key to many of
the earlier lower bounds on quantum communication complexity as well
as to ours. It is easily proven by induction on~$\ell$.

\begin{lemma}[see Yao \cite{yao:qcircuit} and Kremer \cite{kremer:thesis}]\label{lemkremer}
The final state of an $\ell$-qubit protocol on input\/~$(x,y)$ can be written as
\[
\sum_{i\in\{0,1\}^\ell}\ket{A_i(x)}\ket{i_\ell}\ket{B_i(y)},
\]
where the $A_i(x),B_i(y)$ are vectors (of norm\/ $\leq 1$),
and $i_\ell$ denotes the last bit of the $\ell$-bit string~$i$ (the output bit).
\end{lemma}

The acceptance probability~$P(x,y)$ of the protocol is
the squared norm of the part of the final state that has $i_{\ell}=1$.
Letting $a_{ij}$ be the $2^n$-dimensional complex column vector
with the inner products~$\inp{A_i(x)}{A_j(x)}$ as entries
and $b_{ij}$ the $2^n$-dimensional column vector with
entries~$\inp{B_i(y)}{B_j(y)}$,
we can write $P$ (viewed as a $2^n\times 2^n$ matrix) as the sum
$\sum_{i,j:i_{\ell}=j_{\ell}=1}a_{ij}b_{ij}^T$ of $2^{2\ell-2}$ matrices,
each of rank at most~1, so the rank of~$P$ is at most~$2^{2\ell-2}$.
For example, for exact protocols this gives immediately that
$\ell\geq\frac{1}{2}\log {\mbox{\it rank}}(f)+1$, and for nondeterministic
protocols $\ell\geq\frac{1}{2}\log \nrank(f)+1$.

Below we show how we can get rid of the 
factor~$\frac{1}{2}$ in the nondeterministic case and show that
the lower bound of $\log \nrank(f)+1$ is actually optimal.
The lower bound part of the proof relies on the following technical lemma.

\begin{lemma}\label{lemvectorsnrank}
If there exist two families of vectors\/
$\{A_1(x),\ldots,A_m(x)\}\subseteq\mathbb{C}^d$
and\/ $\{B_1(y),\ldots,B_m(y)\}\subseteq\mathbb{C}^d$ such that, for all
$x\in\{0,1\}^n$ and $y\in\{0,1\}^n$, we have
\[
\sum_{i=1}^m A_i(x)\otimes B_i(y)= 0 \mbox{ iff } f(x,y)=0,
\]
then $\nrank(f)\leq m$.
\end{lemma}

\begin{proof}
Assume there exist two such families of vectors.
Let $A_i(x)_j$ denote the $j$th entry of vector~$A_i(x)$,
and similarly let $B_i(y)_k$ denote the $k$th entry of vector~$B_i(y)$.
We use pairs $(j,k)\in\{1,\ldots,d\}^2$ to index entries
of vectors in the $d^2$-dimensional tensor space.
Note that 
\begin{quote}
if $f(x,y)=0$, then $\sum_{i=1}^m A_i(x)_jB_i(y)_k=0$ for all $(j,k)$, and \\
if $f(x,y)=1$, then $\sum_{i=1}^m A_i(x)_jB_i(y)_k\neq 0$ for some $(j,k)$.
\end{quote}
As a first step, we want to replace the vectors $A_i(x)$ and~$B_i(y)$
by numbers $a_i(x)$ and~$b_i(y)$ that have similar properties.
We use the probabilistic method to show that this can be done.

Let $I$ be an arbitrary set of $2^{2n+1}$ numbers.
Choose coefficients $\alpha_1,\ldots,\alpha_d$ and $\beta_1,\ldots,\beta_d$,
each coefficient picked uniformly at random from~$I$.
For every~$x$ define
$a_i(x)=\sum_{j=1}^d\alpha_j A_i(x)_j$,
and for every~$y$ define $b_i(y)=\sum_{k=1}^d\beta_k B_i(y)_k$.
Consider the number
\[
v(x,y)=\sum_{i=1}^m a_i(x)b_i(y)=
\sum_{j,k=1}^d\alpha_j\beta_k\left(\sum_{i=1}^m A_i(x)_jB_i(y)_k\right).
\]
If $f(x,y)=0$, then $v(x,y)=0$ for all choices of the $\alpha_j,\beta_k$.

Now consider some $(x,y)$ with $f(x,y)=1$.  There is a 
pair~$(j',k')$ for which $\sum_{i=1}^m A_i(x)_{j'}B_i(y)_{k'}\neq 0$.
We want to prove that $v(x,y)=0$ happens only with very small probability.
In order to do this, fix the random choices of all
$\alpha_j$, $j\neq j'$, and $\beta_k$,~$k\neq\nobreak k'$,
and view $v(x,y)$ as a function of the two remaining not-yet-chosen
coefficients $\alpha=\alpha_{j'}$ and $\beta=\beta_{k'}$,
\[
v(x,y)=c_0\alpha\beta+c_1\alpha+c_2\beta+c_3.
\]
Here we know that
$c_0=\sum_{i=1}^m A_i(x)_{j'}B_i(y)_{k'}\neq 0$.
There is at most one value of~$\alpha$ for which $c_0\alpha+c_2=0$.
All other values of~$\alpha$ turn $v(x,y)$ into a linear equation
in~$\beta$, so for those $\alpha$ there is at most one choice
of~$\beta$ that gives $v(x,y)=0$.  Hence out of the 
$(2^{2n+1})^2$~different ways of choosing $(\alpha,\beta)$, at most
$2^{2n+1}+(2^{2n+1}-1)\cdot 1<2^{2n+2}$ choices give $v(x,y)=0$.
Therefore,
\[
\Pr[v(x,y)=0]<\frac{2^{2n+2}}{(2^{2n+1})^2}=2^{-2n}.
\]
Using the union bound, we now have
\begin{multline*}
\Pr\left[\mbox{there is an }(x,y)
           \in f^{-1}(1)\mbox{ for which }v(x,y)=0\right] \\
\leq \sum_{(x,y)\in f^{-1}(1)}\Pr[v(x,y)=0]< 2^{2n}\cdot 2^{-2n}=1.
\end{multline*}
This probability is strictly less than~1, so there
exist sets $\{a_1(x),\ldots,a_m(x)\}$ and $\{b_1(y),\ldots,b_m(y)\}$
that make $v(x,y)\neq 0$ for every $(x,y)\in f^{-1}(1)$.
We thus have that
\[
\sum_{i=1}^m a_i(x)b_i(y)=0 \mbox{ iff } f(x,y)=0.
\]
View the $a_i$ and~$b_i$ as $2^n$-dimensional vectors,
let $A$ be the $2^n\times m$ matrix having the $a_i$ as columns,
and let $B$ be the $m\times 2^n$ matrix having the $b_i$ as rows.
Then $(AB)_{xy}=\sum_{i=1}^m a_i(x)b_i(y)$, which is 0 iff $f(x,y)=0$.
Thus $AB$ is a nondeterministic matrix for~$f$,
and $\nrank(f)\leq {\mbox{\it rank}}(AB)\leq {\mbox{\it rank}}(A)\leq m$.\qquad\end{proof}

Lemma~\ref{lemvectorsnrank} allows us to prove the following tight characterization.

\begin{theorem}\label{thNQ=nrank}
$\NQcc(f)=\ceil{\log \nrank(f)}+1$.
\end{theorem}

\begin{proof}
{\it Upper bound}.
Let $r=\nrank(f)$, and let $M$ be a rank-$r$
nondeterministic matrix for~$f$.  Let $M^T=U\Sigma V$ be the
singular value decomposition of the transpose of~$M$~\cite{horn&johnson:ma},
so $U$ and~$V$ are unitary, and $\Sigma$ is a diagonal matrix
whose first $r$~diagonal entries are positive real numbers and
whose other diagonal entries are 0.
Below we describe a one-round nondeterministic protocol for~$f$,
using $\ceil{\log r}+1$ qubits.

First, Alice prepares the state $\ket{\phi_x}=c_x\Sigma V\ket{x}$,
where $c_x>0$ is a normalizing real number that depends on~$x$.
Because only the first $r$~diagonal entries of~$\Sigma$ are nonzero,
only the first $r$~amplitudes of~$\ket{\phi_x}$ are nonzero,
so $\ket{\phi_x}$ can be compressed into $\ceil{\log r}$~qubits.
Alice sends these qubits to Bob.  Bob then applies $U$ to~$\ket{\phi_x}$
and measures the resulting state.
If he observes $\ket{y}$, then he puts 1 on the channel, and otherwise
he puts 0 there.
The acceptance probability of this protocol is
\[
P(x,y) = |\bra{y}U\ket{\phi_x}|^2
       = c_x^2|\bra{y}U\Sigma V\ket{x}|^2
       = c_x^2|M^T_{yx}|^2
       = c_x^2|M_{xy}|^2.
\]
Since $M_{xy}$ is nonzero iff $f(x,y)=1$,
$P(x,y)$ will be positive iff $f(x,y)=1$.
Thus we have a nondeterministic quantum protocol for~$f$
with $\ceil{\log r}+1$ qubits of communication.

{\it Lower bound}.
Consider a nondeterministic $\ell$-qubit protocol for~$f$.
By Lemma~\ref{lemkremer}, its final state on input~$(x,y)$ can be written as
\[
\sum_{i\in\{0,1\}^\ell}\ket{A_i(x)}\ket{i_\ell}\ket{B_i(y)}.
\]
Without loss of generality, we assume the vectors $A_i(x)$ and~$B_i(y)$
all have the same dimension~$d$.  Let $S=\{i\in\{0,1\}^\ell \mid i_{\ell}=1\}$,
and consider the part of the state that corresponds to output 1
(we drop the $i_{\ell}=1$ and the $\ket{\cdot}$-notation here),
\[
\phi(x,y)=\sum_{i\in S}A_i(x)\otimes B_i(y).
\]
Because the protocol has acceptance probability~0 iff $f(x,y)=0$,
this vector~$\phi(x,y)$ will be the zero vector iff $f(x,y)=0$.
The previous lemma gives $\nrank(f)\leq |S|=2^{\ell-1}$;
hence $\log(\nrank(f))+1\leq \NQcc(f)$.\qquad\end{proof}

Note that any nondeterministic matrix for the equality function
has nonzeros on its diagonal and zeros off-diagonal
and hence has full rank.
Thus we obtain $\NQcc(\EQ)=n+1$.
Similarly, a nondeterministic matrix for disjointness has
full rank, because reversing the ordering of the columns
in~$M_f$ gives an upper triangular
matrix with nonzero elements on the diagonal.
This gives tight bounds for the nondeterministic as well
as for the exact setting, neither of which was known prior to this work.

\begin{corollary}
$\Qcc_E(\EQ)=\NQcc(\EQ)=n+1$ and
$\Qcc_E(\DISJ)=\NQcc(\DISJ)\break=n+1$.
\end{corollary}

\subsection{Quantum-classical separation}

To repeat,
classically we have $\Ncc(f)=\ceil{\log \Cov^1(f)}+1$, and quantumly we have
$\NQcc(f)=\ceil{\log \nrank(f)}+1$. We now give a total function~$f$ with
an exponential gap between $\Ncc(f)$~and~$\NQcc(f)$.
For $n>1$, define $f$ by
\[
f(x,y)=1 \ {\mbox {\rm iff}} \ |x\wedge y|\neq 1.
\]
We first show that the quantum complexity~$\NQcc(f)$ is low.

\begin{theorem}
For the above $f$, we have $\NQcc(f)\leq \ceil{\log(n+1)}+1$.
\end{theorem}

\begin{proof}
By Theorem~\ref{thNQ=nrank}, it suffices to prove $\nrank(f)\leq n+1$.
We will derive a low-rank nondeterministic matrix from the polynomial~$p$
of~(\ref{eqndetpoly}), using a technique from~\cite{nisan&wigderson:rank}.
Let $M_i$ be the matrix defined by $M_i(x,y)=1$ if $x_i=y_i=1$
and by $M_i(x,y)=0$ otherwise. Notice that $M_i$ has rank~1.
Define a $2^n\times 2^n$ matrix~$M$ by
\[
M(x,y)=\left(\sum_{i=1}^n M_i(x,y)\right)-1.
\]
Note that $M(x,y)=p(x\wedge y)$.
Since $p$ is a nondeterministic polynomial for the function which is 1 iff its
input does not have weight~1, it can be seen that $M$ is a nondeterministic
matrix for~$f$. Because $M$ is the sum of $n+1$ rank-1
matrices, $M$~itself has rank at most $n+1$.\qquad\end{proof}

Now we show that the classical $\Ncc(f)$ is high (both for $f$ and its complement).

\begin{theorem}
For the above $f$, we have $\Ncc(f)\in\Omega(n)$ and $\Ncc(\overline{f})\geq n-1$.
\end{theorem}

\begin{proof}
Let $R_1,\ldots,R_k$ be a minimal 1-cover for~$f$.
We use the following result
from \cite[Example~3.22 and section~4.6]{kushilevitz&nisan:cc},
which is essentially due to Razborov~\cite{razborov:disj}.
\begin{quote}
There exist sets $A,B\subseteq\01^n\times\01^n$ and a probability
distribution $\mu:\01^n\times\01^n\rightarrow[0,1]$ such that
all $(x,y)\in A$ have $|x\wedge y|=0$,
all $(x,y)\in B$ have $|x\wedge y|=1$,
$\mu(A)=3/4$,
and there are $\alpha,\delta>0$ (independent of~$n$)
such that for all rectangles~$R$,
$\mu(R\cap B)\geq\alpha\cdot \mu(R\cap A)-2^{-\delta n}$.
\end{quote}
Since the $R_i$ are 1-rectangles, they cannot contain elements from~$B$.
Hence $\mu(R_i\cap B)= 0$ and $\mu(R_i\cap A)\leq 2^{-\delta n}/\alpha$.
However, since all elements of~$A$ are covered by the $R_i$, we have
\[
\frac{3}{4}=\mu(A)=\mu\left(\bigcup_{i=1}^k(R_i\cap A)\right)\leq
\sum_{i=1}^k\mu(R_i\cap A)\leq k\cdot \frac{2^{-\delta n}}{\alpha}.
\]
Therefore, $\Ncc(f)=\ceil{\log k}+1\geq\delta n+\log(3\alpha/4)$.

For the lower bound on $\Ncc(\overline{f})$, consider the set
$S=\{(x,y) \mid x_1=y_1=1$, $x_i=\overline{y_i}$ for $i>1\}$.
This $S$ contains $2^{n-1}$~elements, all of which are 1-inputs for $\overline{f}$.
Note that if $(x,y)$ and~$(x',y')$ are two elements from~$S$,
then $|x\wedge y'|>1$ or $|x'\wedge y|>1$,
so a 1-rectangle for~$\overline{f}$ can contain at most one element of~$S$.
This shows that a minimal 1-cover for~$\overline{f}$ requires
at least $2^{n-1}$~rectangles and $\Ncc(\overline{f})\geq n-1$.\qquad\end{proof}

Another quantum-classical separation was obtained earlier by
Massar et~al.~\cite{mbcc:siment}. We include it for the sake of completeness.
It shows that the nondeterministic
complexity of the {\it non\/}equality problem is extremely low,
in sharp contrast to the equality problem itself.

\begin{theorem}[see \cite{mbcc:siment}]\label{thndetcleve}
For the nonequality problem on $n$~bits, 
$\NQcc(\NE)= 2$ versus $\Ncc(\NE)=\log n+1$.
\end{theorem}

\begin{proof}
$\Ncc(\NE)=\log n+1$ is well known (see \cite[Example~2.5]{kushilevitz&nisan:cc}).
Below we give the protocol for $\NE$ from \cite{mbcc:siment}.

Viewing her input~$x$ as a number $\in[0,2^n-1]$,
Alice rotates a $\ket{0}$-qubit over an angle~$x\pi/2^n$, obtaining a qubit
$\cos(x\pi/2^n)\ket{0}+\sin(x\pi/2^n)\ket{1}$ which she sends to Bob.
Bob rotates the qubit back over an angle $y\pi/2^n$, obtaining
$\cos((x-\nobreak y)\pi/2^n)\ket{0}+\sin((x-\nobreak y)\pi/2^n)\ket{1}$.
Bob now measures the qubit and sends back the observed bit.
If $x=y$, then $\sin((x-y)\pi/2^n)=0$, so Bob will always send 0.
If $x\neq y$, then $\sin((x-y)\pi/2^n)\neq 0$,
so Bob will send 1 with positive probability.\qquad\end{proof}

In another direction, Klauck~\cite{klauck:qcclower}
showed that $\NQcc(f)$ is in general incomparable to bounded-error
quantum communication complexity: the latter may be exponentially larger
or smaller, depending on~$f$.

\section{Future work}

One of the main reasons for the usefulness of nondeterministic query and
communication complexities in the classical case is the tight relation
of these complexities with deterministic complexity.

In the query complexity (decision tree) setting, we have the well-known bound
\[
\max\{\Nqc(f), \Nqc(\overline{f})\}\leq \Dqc(f)\leq \Nqc(f)\Nqc(\overline{f}).
\]
We conjecture that something similar holds in the quantum case:
\[
\max\left\{\NQqc(f),\NQqc(\overline{f})\right\}
  \leq \Qqc_E(f)\leq \Dqc(f)
\stackrel{?}{\leq}O(\NQqc(f)\NQqc(\overline{f})).
\]
The $?$-part is open and
ties in with tightly embedding $\NQqc(f)$ and~$\ndeg(f)$
into the web of known relations between various complexity measures
(section~\ref{ssecembedndeg}).
This conjecture implies, for instance, $\Dqc(f)\in O(\deg(f)^2)$,
which would be close to optimal~\cite{nisan&szegedy:degree}.
Similarly, it would imply $\Dqc(f)\in O(\Qqc_0(f)^2)$,
which would be close to optimal as well~\cite{bcwz:qerror}.
In both cases, the currently best  relation
has a fourth power instead of a square.

Similarly, for communication complexity, the following is
known \cite[section~2.11]{kushilevitz&nisan:cc}:
\[
\max\{\Ncc(f), \Ncc(\overline{f})\}\leq \Dcc(f)\leq O(\Ncc(f)\Ncc(\overline{f})).
\]
An analogous result might be true in the quantum setting,
but we have been unable to prove it.
So far, the best result in this direction is Klauck's observation that
$\Dcc(f)=O(\Ncc(f)\NQcc(\overline{f}))$ \cite[Theorem~1]{klauck:qccsurvey}.

{\Appendix
\section{Comparison with alternative definitions}
As mentioned in the introduction, three different definitions of
nondeterministic quantum complexity are possible.
We may consider the complexity of quantum algorithms that 
\begin{enumerate}
\item output 1 iff given an appropriate {\it classical\/} certificate
(and such certificates must exist iff $f(x)=1$),
\item output 1 iff given an appropriate {\it quantum\/} certificate
(and such certificates must exist iff $f(x)=1$), or
\item  output 1 with positive probability iff $f(x)=1$.
\end{enumerate}
The third definition is the one we adopted for this paper.
Clearly definition~2 is at least as strong as definition~1
in the sense that the complexity of a function according
to definition~2 will be less than or equal to the complexity
according to definition~1. In fact, in the setting of query complexity,
these two definitions are equivalent, because without loss of generality
the certificate can be taken to be the purported input.
See Aaronson~\cite{aaronson:qcert} for some recent results about
``quantum certificate (query) complexity.''

Here we show that definition~3 is at least as strong as definition~2.
We give the proof for the query complexity setting, but the same proof
can be modified to work for communication complexity and other nonuniform
settings as well. We then give an example in which the query complexity
according to definition~3 is much less than according to definition~2.
This shows that our $\NQqc(f)$ is in fact the most powerful
definition of nondeterministic quantum query complexity.

We formalize definition~2 as follows.
A {\it $T$-query quantum verifier\/} for $f$ is a $T$-query quantum
algorithm~$V$ together with a set~$\cal C$ of $m$-qubit states,
such that for all $x\in\01^n$ we have
(1)~if $f(x)=1$, then there is a $\ket{\phi_x}\in{\cal C}$
such that $V_x\ket{\phi_x}$ has acceptance probability~1; and
(2)~if $f(x)=0$, then $V_x\ket{\phi}$ has acceptance
probability~0 for every $\ket{\phi}\in{\cal C}$.
Informally, the set $\cal C$ contains all possible certificates:
(1)~for every 1-input, there is a verifiable 1-certificate in~$\cal C$;
and (2)~for 0-inputs, there are not any.
We do not put any constraints on~$\cal C$.
However, note that the definition implies that if $f(x)=0$ for some~$x$,
then $\cal C$ cannot contain {\it all} $m$-qubit states;
otherwise, $\ket{\phi_x}=V_x^{-1}\ket{1\vec{0}}$ would be
a 1-certificate in~$\cal C$ even for $x$ with $f(x)=0$.

We now prove that a $T$-query quantum verifier can be turned into a
$T$-query nondeterministic quantum algorithm according to our third definition.
This shows that the third definition is at least as powerful as the second.
In fact, this even holds if we replace the acceptance probability~1
in clause~(1) of the definition of a quantum verifier by just positive
acceptance probability---in this case, both definitions are equivalent.

\begin{theorem}\label{th2ndetdef}
If there is a $T$-query quantum verifier~$V$ for~$f$,
then $\NQqc(f)\leq T$.
\end{theorem}

\begin{proof}
The verifier~$V$ and the associated set~$\cal C$ satisfy the following:
\begin{enumerate}
\item If $f(x)=1$, then there is a $\ket{\phi_x}\in{\cal C}$
such that $V_x\ket{\phi_x}$ has acceptance probability~1.
\item If $f(x)=0$, then $V_x\ket{\phi}$ has acceptance probability~0 for all
$\ket{\phi}\in{\cal C}$.
\end{enumerate}
Let $X_1=\{z\mid f(z)=1\}$.
For each $z\in X_1$, choose one specific 1-certificate $\ket{\phi_z}\in{\cal C}$.
Now let us consider some input~$x$ and see what happens if we run $V_x\otimes I$
(where $I$ is the $2^n\times 2^n$ identity operation) on the $m+n$-qubit state
\[
\ket{\phi}=\frac{1}{\sqrt{|X_1|}}\sum_{z\in X_1}\ket{\phi_z}\ket{z}.
\]
$V_x$ acts on only the first $m$ qubits of~$\ket{\phi}$;
the $\ket{z}$-part remains unaffected.
Therefore, running $V_x\otimes I$ on~$\ket{\phi}$ gives the same
acceptance probabilities as when we first randomly choose
some $z\in X_1$ and then apply $V_x$ to~$\ket{\phi_z}$.
In the case when $f(x)=0$, this $V_x\ket{\phi_z}$ will have acceptance probability~0,
so $(V_x\otimes I)\ket{\phi}$ will have acceptance probability~0 as well.
In the case when the input~$x$ is such that $f(x)=1$,
the specific certificate~$\ket{\phi_z}$ that we chose for
this $x$ will satisfy that $V_x\ket{\phi_x}$ has acceptance probability~1.
However, then $(V_x\otimes\nobreak I)\ket{\phi}$ has acceptance probability at least $1/|X_1|>0$.
Accordingly, $(V_x\otimes\nobreak I)\ket{\phi}$ has positive acceptance probability iff $f(x)=1$.
By prefixing $V_x\otimes I$ with a unitary transformation that maps $\ket{\vec{0}}$
(of $m+n$ qubits) to~$\ket{\phi}$, we have constructed a nondeterministic
quantum algorithm for~$f$ with $T$~queries.\qquad\end{proof}

The above proof shows that our definition of~$\NQqc(f)$ is at least as
strong as the certificate-verifier definition.  Could it be that both
definitions are in fact equivalent (i.e., yield the same complexity)?
The function we used in section~\ref{ssecnqquerysep} shows that
this is not the case. Consider again
\[
f(x)=1 \ {\mbox {\rm iff}} \ |x|\neq 1.
\]
It satisfies $\NQqc(f)=1$. On the other hand, if we take a $T$-query
verifier for~$f$ and fix the certificate for the all-0 input,
we obtain a $T$-query algorithm that always outputs~1 on the
all-0 input and that outputs~0 on all inputs of Hamming weight~1.
The quantum search lower bounds~\cite{bbbv:str&weak,bbcmw:polynomials}
immediately imply $T=\Omega(\sqrt{n})$.
This shows that our definition of $\NQqc(f)$ is strictly
more powerful than the certificate-verifying one.}

\section*{Acknowledgments}
Many thanks to Peter H\o yer for stimulating discussions and for his
permission to include the nondeterministic parts of our joint
paper~\cite{hoyer&wolf:disjeq} in this paper. I also thank
Scott Aaronson, Harry Buhrman, Richard Cleve,
Wim van~Dam, Lance Fortnow, Hartmut Klauck, Peter Shor, and John Watrous
for various helpful discussions, comments, and pointers to the literature.
Thanks to the anonymous referees for suggesting some improvements.



\end{document}